\documentclass[%
 reprint,
 amsmath,amssymb,
 prl,
]{revtex4-2}

\usepackage{graphicx}
\usepackage{dcolumn}
\usepackage{bm}

\begin{document}
\title{Emergent Detailed Balance in Human Mobility under Temporal Coarse-Graining}

\author{Lei Dong}
\thanks{leidong@pku.edu.cn}

\affiliation{%
 Institute of Remote Sensing and Geographical Information Systems, School of Earth and Space Sciences, Peking University, Beijing 100087, China
}%

\date{\today}

\begin{abstract}
A fundamental question in nonequilibrium statistical physics is whether effective equilibrium behavior can emerge at coarse-grained scales in strongly driven systems. Here, we investigate this question in the context of human mobility by analyzing five years of intercity flow data covering millions of travelers. While short-term flows are highly asymmetric, temporal coarse-graining reveals that over half of all city pairs converge toward effective flow balance, with normalized directional imbalance decaying as a power law. The remaining pairs either exhibit persistent drift-dominated currents or a crossover between these two extremes. A stochastic model decomposing mobility into directional drift and correlated fluctuations quantitatively captures the coexistence of all three regimes. Directly measured variance scaling of the fluctuation process confirms near-diffusive behavior with regime-dependent deviations. These results demonstrate that large-scale mobility networks exhibit a scale-dependent transition from broken to restored flow symmetry, with direct implications for modeling transport and spreading dynamics.
\end{abstract}

\maketitle

{\it Introduction.---}A central question in nonequilibrium statistical physics is whether and how effective equilibrium behavior can emerge from coarse-graining of strongly driven dynamics. In Markovian systems, detailed balance --- the condition $\pi_i P_{ij} = \pi_j P_{ji}$ relating stationary distributions to transition rates --- is the hallmark of thermodynamic equilibrium \cite{risken1989fokker,seifert2012stochastic}. Its violation signals irreversibility and is intimately linked to entropy production \cite{gnesotto2018broken,platini2011measure,battle2016broken,gonzalez2019experimental}. Recent theoretical work has shown that the detection and magnitude of detailed-balance violation depend sensitively on the level of coarse-graining, whether spatial or temporal \cite{egolf2000equilibrium,dieball2022mathematical,teza2020exact,cocconi2022scaling}. Yet empirical demonstrations of scale-dependent transitions between nonequilibrium and effective equilibrium behavior in real-world complex systems remain scarce.

Human mobility provides a natural setting to probe this question. Intercity population flows are shaped by economic gradients, infrastructure constraints, and policy interventions \cite{mcfadden1973conditional,brockmann2006scaling,bassolas2019hierarchical,tian2020investigation,barbosa2018human}, generating directed fluxes characteristic of nonequilibrium transport. Yet many city pairs sustain intense bidirectional exchange through commuting, tourism, and recurrent travel \cite{mazzoli2019field,schlapfer2021universal}, raising a fundamental question: do the directional asymmetries observed in mobility data represent persistent nonequilibrium currents, or transient fluctuations that vanish under temporal aggregation?

We emphasize that in this Letter, we use ``detailed balance'' in an effective, flow-level sense: the condition that forward and reverse population fluxes between city pairs are equal. This is a necessary consequence of, but not identical to, the strict Markovian detailed balance condition $\pi_i P_{ij} = \pi_j P_{ji}$ defined on transition probabilities weighted by stationary populations. In networks where the stationary population distribution $\pi_i$ is approximately constant over the timescales considered, as is the case for intercity flows on weekly-to-annual scales, the two conditions become equivalent. We discuss this equivalence explicitly in the Supplemental Material (SM).

Distinguishing transient fluctuations from persistent currents is challenging, as it requires mobility data spanning sufficiently long timescales together with a large number of city pairs to resolve heterogeneity across links. This distinction has direct implications: if flows approach balance under coarse-graining, they admit an effective equilibrium description \cite{brockmann2013hidden,zhang2020investigating}; if directional biases persist, the system retains genuine nonequilibrium currents \cite{derrida2007non}. Such persistent currents can bias spreading dynamics on spatial networks \cite{balcan2011phase,belik2011natural,jia2020population,davis2020phase} and influence spatial pattern formation processes including urban growth and regional development \cite{verbavatz2020growth,reia2022spatial,yakubo2014superlinear}.

In this Letter, we analyze five years of intercity mobility data and demonstrate that flow balance is an emergent, scale-dependent property. A stochastic model with directly measured parameters, including the diffusion exponent of mobility fluctuations, quantitatively captures the coexistence of three distinct transport regimes.

\begin{figure*}[ht]
    \centering
    \includegraphics[width = 0.9\textwidth]{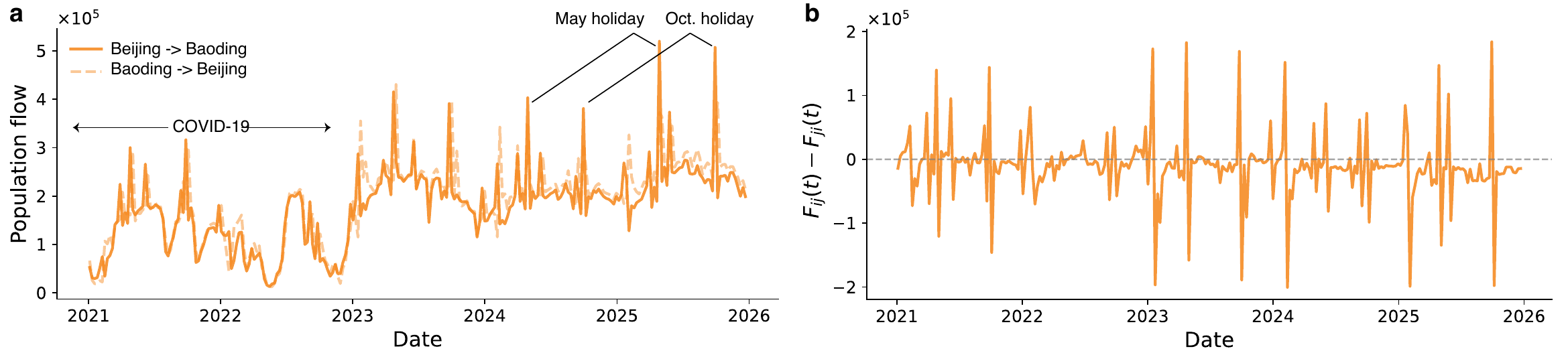}
    \caption{Weekly mobility flows between Beijing and Baoding. (a) The bidirectional flows exhibit similar temporal patterns, with clear disruptions during the COVID-19 and holiday peaks after 2023. (b) The net flow reveals strong fluctuations, frequently reaching 30--40\% of the total bidirectional volume.}
    \label{fig1}
\end{figure*}

{\it Data.---}We analyze five years (2021--2025) of daily intercity mobility flows from the Baidu Qianxi project, which infers movements from mobile phone geolocation data \cite{baidu}. The data cover nearly all prefecture-level cities in China, providing a comprehensive view of large-scale population transport. To smooth out weekly periodicities, we aggregate the raw data into weekly flows and restrict our analysis to city pairs with an average weekly flow of at least 100 individuals, ensuring statistical reliability. This yields approximately $1.2 \times 10^4$ city pairs for subsequent analysis (see SM for details).

For each city pair $(i,j)$, let $F_{ij}(t)$ denote the number of individuals traveling from $i$ to $j$ during week $t$. If flow balance holds, we would observe $F_{ij}(t)=F_{ji}(t)$ for all pairs. In practice, short-term flows are highly asymmetric. Figure~\ref{fig1} illustrates this point for Beijing and Baoding, two neighboring cities with intense exchange. While the bidirectional flows exhibit similar temporal trends, the net imbalance $F_{ij}(t)-F_{ji}(t)$ is substantial, frequently reaching 30--40\% of the total bidirectional traffic (Fig.~\ref{fig1}b). This motivates a scale-dependent measure of imbalance.

{\it Imbalance measure.---}To quantify the directional asymmetry, we define the instantaneous net flow $X_{ij}(t)=F_{ij}(t)-F_{ji}(t)$ and the total bidirectional traffic $Y_{ij}(t)=F_{ij}(t)+F_{ji}(t)$. Aggregating over a window of length $\tau$ yields the cumulative imbalance $D_{ij}^{(\tau)}(t)=\sum_{s=t}^{t+\tau-1}X_{ij}(s)$ and total traffic $S_{ij}^{(\tau)}(t)=\sum_{s=t}^{t+\tau-1}Y_{ij}(s)$. The normalized imbalance
\begin{equation}
R_{ij}^{(\tau)}(t)=\frac{|D_{ij}^{(\tau)}(t)|}{S_{ij}^{(\tau)}(t)},
\end{equation}
measures the relative violation of flow balance between city pairs at temporal scale $\tau$.

\begin{figure}[ht]
    \centering
    \includegraphics[width =.5\textwidth]{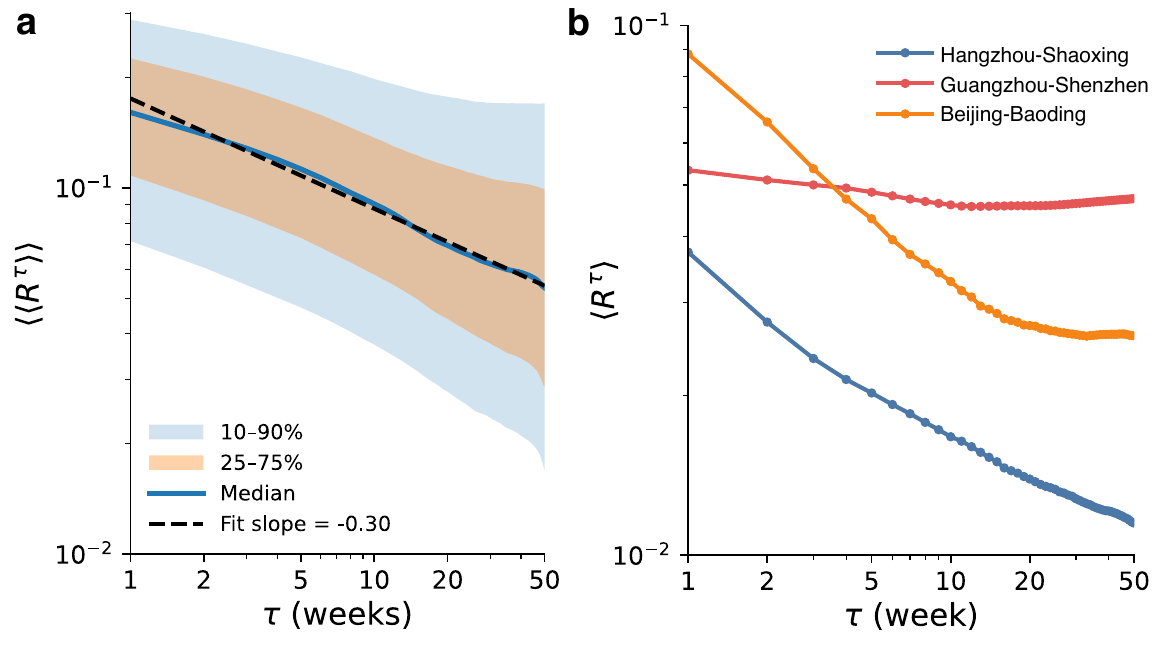}
    \caption{(a) Ensemble-averaged normalized imbalance $\langle\langle R^{(\tau)} \rangle \rangle$ as a function of aggregation window $\tau$. The dashed line indicates a power-law fit with exponent $-0.30$. (b) Normalized imbalance trajectories for a representative sample of city pairs, illustrating heterogeneous behavior.}
    \label{fig2}
\end{figure}

For each city pair, we compute the time-averaged $\langle R_{ij}^{(\tau)}(t) \rangle$ over all starting weeks $t$ in the five-year record. The ensemble average $\langle \langle R^{(\tau)} \rangle \rangle$ across all pairs decreases systematically with $\tau$ from one week to nearly one year, following an approximate power law $\langle  \langle R^{(\tau)} \rangle \rangle \sim \tau^{\alpha}$ with exponent $\alpha \approx -0.30$ (Fig.~\ref{fig2}a). This decay suggests that short-term directional fluctuations tend to partially cancel under coarse-graining.

The ensemble average, however, masks a profound heterogeneity. Figure~\ref{fig2}b shows individual trajectories $R_{ij}^{(\tau)}$ for a representative sample of city pairs: some decay continuously, others saturate to a finite plateau, and yet others exhibit a clear bend between these behaviors. This diversity indicates that the mobility network is not a monolithic nonequilibrium system but a composite of dynamically distinct links.

\begin{figure*}[ht]
    \centering
    \includegraphics[width = 0.9\textwidth]{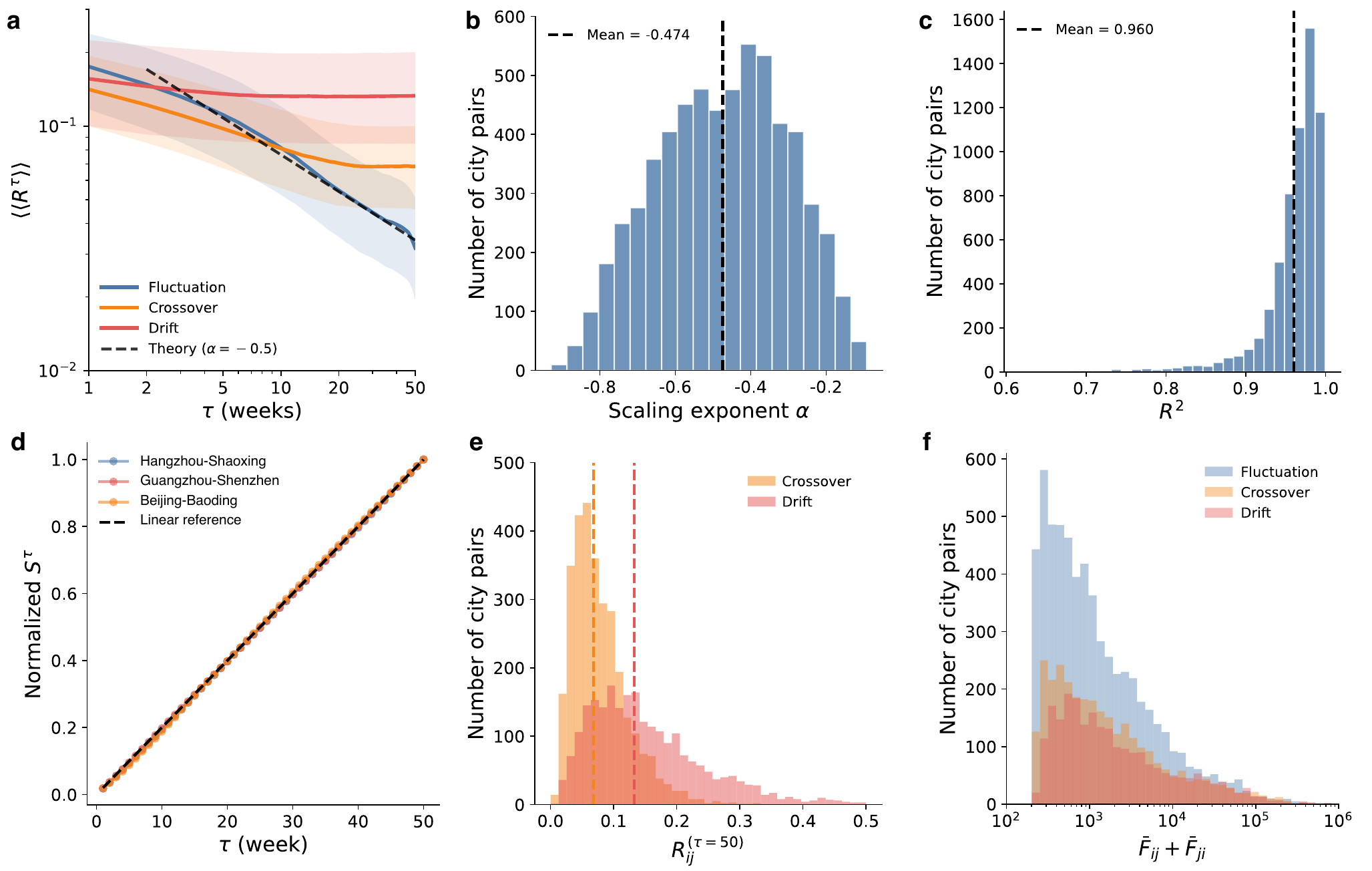}
    \caption{Three empirical regimes of directional imbalance revealed by the scaling behavior. (a) Regime-averaged normalized imbalance $\langle \langle R^{(\tau)} \rangle \rangle$ as a function of the aggregation window $\tau$ for city pairs classified into three regimes: fluctuation-dominated decay (blue), crossover to plateau (orange), and drift-dominated saturation (red). Shaded regions indicate the interquartile range (25--75\%). The dashed line shows the theoretical prediction for uncorrelated fluctuations (slope = -0.5). (b) Distribution of decay exponents for fluctuation-dominated links. (c) Distribution of goodness-of-fit, measured by $R^2$, for the power-law fits. (d) Growth of cumulative bidirectional traffic with aggregation window $\tau$, showing the approximately linear scaling $S_{ij}^{(\tau)} \propto \tau$. Flows are normalized by their maximum value for comparison across city pairs. (e) Distribution of plateau values of $R^{(\tau)}$ for drift-dominated and crossover links. (f) Distribution of mean weekly flow for the three regimes.}
    \label{fig3}
\end{figure*}

{\it Empirical transport regimes.---}
To systematically classify these patterns, we analyze the scaling behavior of each pair by estimating local exponents $\alpha = \frac{d\log R_{ij}^{(\tau)}}{d\log\tau}$ over small-$\tau$ and large-$\tau$ regimes. A city pair is classified as fluctuation-dominated if both slopes satisfy $\alpha < -0.1$; as drift-dominated if $|\alpha| < 0.1$ at both scales; and as crossover if it decays at small $\tau$ but flattens at large $\tau$ (see SM for sensitivity analysis). A small fraction of pairs ($<$5\%) exhibit atypical patterns that do not fall into these three categories and are excluded from the subsequent analysis (see SM). This procedure reveals three dominant transport regimes (Fig.~\ref{fig3}a):

\textit{Fluctuation-dominated decay.} For the majority of city pairs (53\%), $R^{(\tau)}$ decays as a power law across all observed scales, approaching zero at large $\tau$ (blue line, Fig.~\ref{fig3}a). The decay exponents are centered at $\langle \alpha \rangle  = -0.474 \pm 0.17$ (mean $\pm$ std) with high goodness of fit (Fig.~\ref{fig3}b,c). These exponents are close to the $-0.5$ expected for uncorrelated stochastic fluctuations. Here, flow balance is effectively restored under coarse-graining.

\textit{Drift-dominated saturation.} A substantial minority (21\%) display nearly constant imbalances at all aggregation scales (red line, Fig.~\ref{fig3}a), saturating to a finite plateau with median 0.13 (5--95\% range: 0.04--0.33; Fig.~\ref{fig3}e). They represent persistent directional currents that violate flow balance at all observable scales.

\textit{Crossover regime.} The remaining links (26\%) initially decay but flatten beyond a characteristic scale of 20--30 weeks (orange line, Fig.~\ref{fig3}a), reaching a lower median plateau of 0.068 (5--95\% range: 0.024--0.165; Fig.~\ref{fig3}e). They reflect a competition between fluctuations and drift, with the latter becoming dominant at long times.

These classifications are robust: restricting the analysis to the post-COVID period (2023--2025) preserves the three-regime structure and all scaling exponents within statistical uncertainty. Results are also insensitive to the specific thresholds used for regime classification (see SM, Fig.~S1 and Table.~S1).

{\it Stochastic model.---}The coexistence of these three regimes raises a natural question: Can a single theoretical framework explain why some city pairs approach flow balance under temporal coarse-graining while others maintain persistent directional bias? To address this question, we develop a stochastic model that captures the competition between random mobility fluctuations and systematic directional drift \cite{gardiner2009stochastic,kampen2007stochastic}.

Let the directional increment $X_{ij}(t)$ between cities $i,j$ be decomposed as

\begin{equation}
X_{ij}(t) = \mu_{X} + \eta(t),
\label{eq:decomposition}
\end{equation}
where $\mu_{X} = \langle X_{ij}(t) \rangle$ is the mean directional drift and $\eta(t)$ is a zero-mean fluctuating term. The cumulative imbalance over a window of length $\tau$ is

\begin{equation}
D_{ij}^{(\tau)} = \sum_{t=1}^{\tau} X_{ij}(t) = \mu_{X}\tau + \sum_{t=1}^{\tau} \eta(t).
\label{eq:cumulative_decomposition}
\end{equation}

The key physical quantity is the variance scaling of the cumulative fluctuation process. We measure this directly from the data by computing $\mathrm{Var}\left(\sum_{t=1}^{\tau} \eta(t)\right)$ for each city pair after subtracting the estimated drift $\hat{\mu}_X$. The variance grows as $\mathrm{Var}\left(\sum_{t=1}^{\tau} \eta(t)\right) \propto \tau^{\beta}$ \cite{metzler2000random,mandelbrot1968fractional}, where the exponent $\beta$ encodes temporal correlations in the fluctuation process: $\beta = 1$ for uncorrelated (white) noise, $\beta > 1$ for positively correlated (superdiffusive) fluctuations, and $\beta < 1$ for anticorrelated (subdiffusive) behavior.

Across all city pairs, $\langle \beta \rangle = 1.06 \pm 0.35$ (mean $\pm$ std), indicating that fluctuations are on average approximately uncorrelated but with substantial heterogeneity across links (see SM, Fig.~S2). Notably, the exponent varies systematically across regimes: fluctuation-dominated links yield $\langle \beta \rangle = 0.98 \pm 0.33$, consistent with near-white-noise behavior and weak anticorrelations suggestive of mean-reverting dynamics; drift-dominated and crossover links, by contrast, tend toward $\beta > 1$, indicating that the same socioeconomic forces sustaining persistent currents also induce positively correlated fluctuations. For fluctuation-dominated links, the model predicts $\alpha = \beta/2 - 1$, yielding $\alpha \approx -0.51$ for $\beta \approx 0.98$, in good agreement with the measured mean $\alpha = -0.474$ (Fig.~\ref{fig3}b). The residual difference lies well within the width of the exponent distribution.

The typical magnitude of cumulative imbalance involves two terms. For the expectation of the absolute cumulative imbalance, we note that when $\mu_X = 0$, $\langle |D^{(\tau)}| \rangle \propto \sigma \tau^{\beta/2}$ by the scaling of the standard deviation. When $\mu_X \neq 0$ and $|\mu_X|\tau \gg \sigma\tau^{\beta/2}$, $|D^{(\tau)}| \approx |\mu_X|\tau$. In general, the two contributions combine as

\begin{equation}
\langle |D_{ij}^{(\tau)}| \rangle \sim \sqrt{\mu_X^2 \tau^2 + \sigma^2 \tau^{\beta}} \,,
\label{eq:competition}
\end{equation}
which interpolates smoothly between the fluctuation-dominated ($\sigma\tau^{\beta/2}$) and drift-dominated ($|\mu_X|\tau$) limits.

Since the total bidirectional traffic grows linearly (Fig.~\ref{fig3}d and Fig.~S3), $S_{ij}^{(\tau)} \approx \mu_Y\tau$ with $\mu_Y = \langle Y_{ij}(t)\rangle$, the normalized imbalance becomes

\begin{equation}
R_{ij}^{(\tau)} \approx \frac{1}{\mu_Y}\sqrt{\mu_X^2 + \sigma^2 \tau^{\beta-2}}.
\label{eq:R_tau_final}
\end{equation}

Equation~(\ref{eq:R_tau_final}) predicts all three regimes depending on the dimensionless ratio $\rho \equiv |\mu_X|/(\sigma \tau_0^{\beta/2-1})$ that compares drift to fluctuation strength at the elementary timescale $\tau_0 = 1$ week:

\textit{Fluctuation-dominated decay} ($\rho \ll 1$). When directional drift is negligible, the normalized imbalance is governed by fluctuations: $R^{(\tau)} \sim \tau^{\beta/2-1}$. For $\beta \approx 0.98$, this yields $R^{(\tau)} \sim \tau^{-0.51}$, in good agreement with the observed exponent distribution (Fig.~\ref{fig3}b).

\textit{Drift-dominated saturation} ($\rho \gg 1$). The normalized imbalance is approximately constant:
\begin{equation}
R_{ij}^{(\tau)} \approx \frac{|\mu_{X}|}{\mu_Y},
\end{equation}
corresponding to the empirical plateau (median $\approx 0.13$; Fig.~\ref{fig3}e). These city pairs sustain persistent directional bias at all observable scales.

\textit{Crossover regime} ($\rho \sim 1$). Fluctuations dominate at short scales ($R^{(\tau)} \sim \tau^{\beta/2-1}$), while drift becomes dominant beyond a crossover scale
\begin{equation}
\tau_c \sim \left(\frac{\sigma}{|\mu_X|}\right)^{2/(2-\beta)},
\end{equation}
which marks where the drift and fluctuation contributions to $R^{(\tau)}$ are equal. Using measured values of $\sigma$, $\mu_X$, and $\beta$ for crossover links, we obtain a median $\tau_c \approx 8$ weeks (see SM). The visual flattening of $R^{(\tau)}$ in Fig.~\ref{fig3}a occurs at $\sim$20--30 weeks, corresponding to $\sim 3\,\tau_c$, which is the scale at which fluctuations become fully subdominant and the plateau is established.

The model also explains why the ensemble-averaged scaling (Fig.~\ref{fig2}a) is shallower than the fluctuation-dominated exponent: it reflects a mixture of links from different regimes.

{\it Entropy production.---}To connect our analysis with the standard nonequilibrium framework, we compute the network-level entropy production rate associated with flow asymmetry. Following \cite{seifert2012stochastic,gnesotto2018broken}, we define
\begin{equation}
\dot{S}^{(\tau)} = \sum_{\langle i,j\rangle} \left(J_{ij}^{(\tau)} - J_{ji}^{(\tau)}\right) \ln \frac{J_{ij}^{(\tau)}}{J_{ji}^{(\tau)}},
\end{equation}
where $J_{ij}^{(\tau)} = \frac{1}{\tau}\sum_{t=1}^{\tau} F_{ij}(t)$ is the time-averaged flux over window $\tau$ and the sum runs over all city pairs with $J_{ij}^{(\tau)}, J_{ji}^{(\tau)} > 0$. Each term is non-negative by the log-sum inequality, so $\dot{S}^{(\tau)} \geq 0$, with equality only when $J_{ij} = J_{ji}$ for all pairs. We find that $\dot{S}^{(\tau)}$ decreases monotonically with $\tau$, with the dominant contribution shifting from fluctuation-dominated links at small $\tau$ to drift-dominated links at large $\tau$, where it saturates to a finite value set by the persistent currents (SM, Fig.~S4). This confirms that the mobility network undergoes a partial, but not complete, restoration of time-reversal symmetry under coarse-graining.

\begin{figure*}[ht]
    \centering
    \includegraphics[width = 0.9\textwidth]{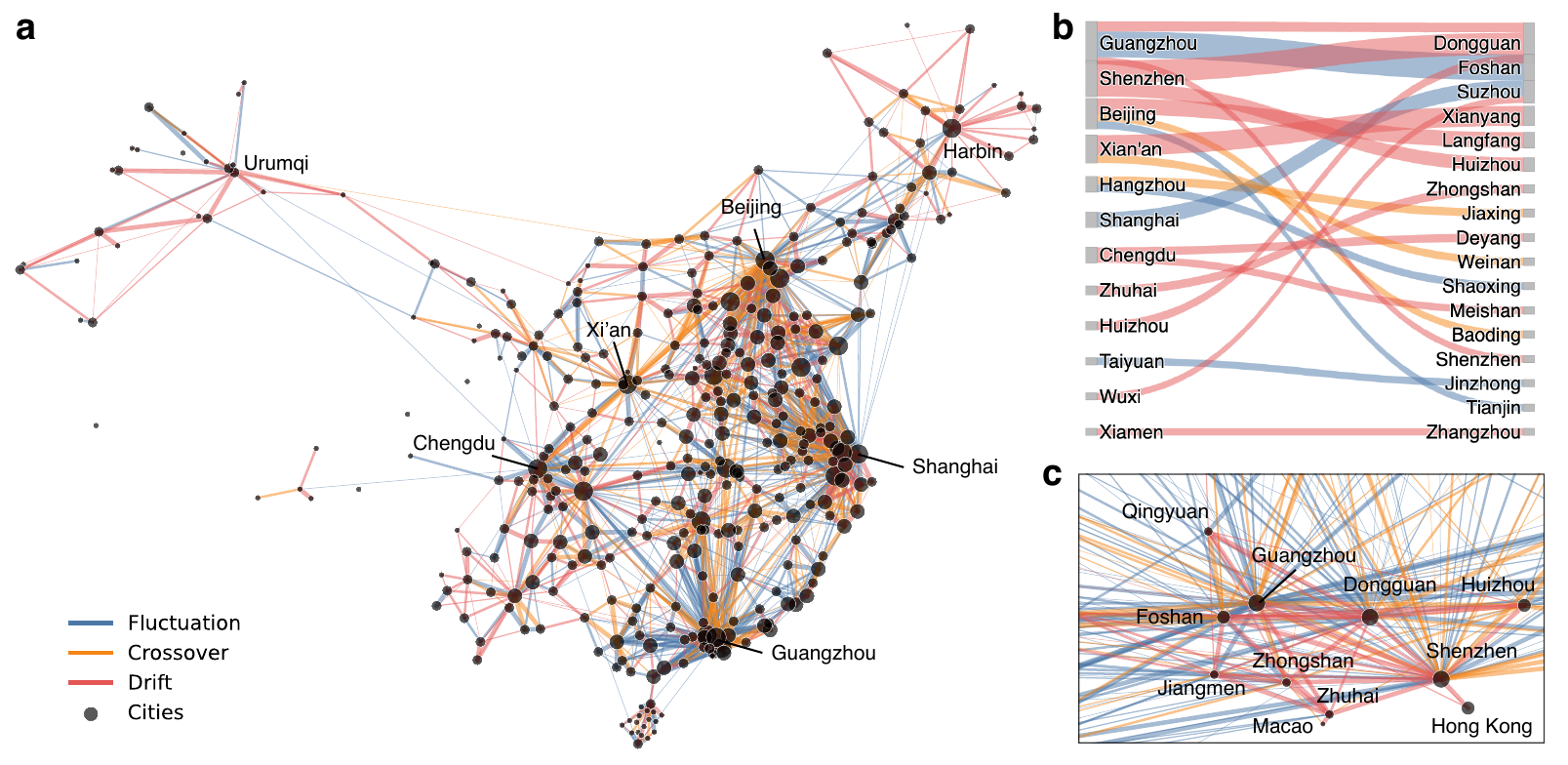}
    \caption{Spatial distribution of mobility regimes. (a) Network map of the 2,000 city pairs with the largest weekly mean flows. Colors indicate the three transport regimes identified in Fig.~\ref{fig3} (fluctuation-dominated, crossover, and drift-dominated). Node size is proportional to city population from the 2020 census. (b) The twenty city pairs with the largest mean flows. (c) The Guangdong Bay Area, corresponding to several high-traffic links highlighted in panel (b).}
    \label{fig4}
\end{figure*}

{\it Spatial organization.---}The three regimes are not simply distinguished by flow volume: links with similar weekly flows can fall into any regime (Fig.~\ref{fig3}f). This indicates that persistent directional imbalance is not simply determined by mobility magnitude.

To further examine the spatial structure of these regimes, Fig.~\ref{fig4} visualizes their distribution on the intercity network. Drift-dominated links are concentrated primarily in northeastern, northwestern, and southwestern China, where many flows point toward regional centers,
suggesting long-lived source--sink relationships.

Among the largest mobility connections (Fig.~\ref{fig4}b), some geographically close and economically integrated city pairs, such as Guangzhou--Foshan, Shanghai--Suzhou, and Beijing--Tianjin, remain nearly balanced after temporal aggregation. In contrast, links such as Shenzhen--Dongguan and Shenzhen--Huizhou exhibit strong directional bias, suggesting population redistribution from the metropolis to surrounding cities.

We caution that persistent directional drift in aggregate flows does not necessarily imply permanent migration. The observed imbalance could also arise from multi-city \emph{circulation flows} (i.e., closed paths involving three or more cities) that produce pairwise asymmetries even without net population transfer. Distinguishing these mechanisms would require trajectory-level data.

{\it Discussion.---} Our results demonstrate that flow balance in human mobility is a scale-dependent property. At short timescales, intercity flows exhibit pronounced directional asymmetries consistent with nonequilibrium transport. Under temporal coarse-graining, a majority of city pairs converge toward effective balance, with normalized imbalance decaying as a power law.

We note that for a single city pair with purely random, uncorrelated fluctuations ($\beta = 1$, $\mu_X = 0$), the decay $R^{(\tau)} \sim \tau^{-1/2}$ follows trivially from the central limit theorem. The nontrivial content of our findings lies in three aspects. First, the empirical observation that over half of all links in a large-scale socioeconomic transport network are well described by this null model is itself surprising, given the strong heterogeneity and external driving. Second, the coexistence of three dynamically distinct regimes --- and their spatial organization --- reveals that the mobility network is not uniformly driven but contains a hidden equilibrium backbone interspersed with genuinely nonequilibrium corridors. Third, the direct measurement of the fluctuation exponent $\beta$ shows that mobility fluctuations are close to, but not exactly, uncorrelated white noise, connecting mobility dynamics to the broader framework of anomalous transport \cite{metzler2000random}.

The relationship between coarse-graining and detailed balance has been studied extensively in the context of \emph{spatial} coarse-graining, where lumping microstates into mesostates generically generates apparent entropy production and can either mask or create the appearance of irreversibility \cite{dieball2022mathematical,teza2020exact,cocconi2022scaling}. Our work addresses the complementary question of \emph{temporal} coarse-graining: aggregating dynamics over longer observation windows. While spatial coarse-graining typically obscures microscopic irreversibility, we show that temporal coarse-graining can \emph{reveal} an underlying effective equilibrium by averaging out transient fluctuations. Our empirical demonstration resonates with the ``equilibrium regained'' phenomenon in driven chaotic systems \cite{egolf2000equilibrium}, extending it to a data-driven, socially relevant context.

Finally, these findings carry practical implications for modeling spreading processes. Many metapopulation frameworks assume symmetric or diffusion-like mobility \cite{pastor2015epidemic,liu2013contagion}; our results show that such approximations are valid only for the fluctuation-dominated subset of the network. Drift-dominated corridors sustain persistent currents that can bias propagation pathways of infectious diseases, information, or other perturbations \cite{granell2018epidemic,winn2023localized}. We note that our analysis is based on a single country (China); validating the generality of these findings across different geographies and transport modes is an important direction for future work.

{\it Acknowledgment.---}We thank Junjie Yang and Junlong Zhang for useful discussions. L.D. acknowledges support from the National Natural Science Foundation of China (Grant No. 42422110).

{\it Contributions.---}L.D. designed the study, analyzed and interpreted the data and wrote the manuscript.

{\it Competing interests.---}The authors declare they have no conflicts of interest.

{\it Data availability.---}The data that support the findings of this article are openly available.

\bibliographystyle{unsrt} 
\bibliography{ref.bib} 

\end{document}